\documentclass[aps,prl,twocolumn,amsmath,superscriptaddress]{revtex4}

\usepackage{graphicx}
\usepackage{color}

\begin{document}

\title{Optical Stark Effect and Dressed Excitonic States in a Mn-doped Quantum Dot}

\author{C. Le Gall}
\author{A. Brunetti}
\author{H. Boukari}
\author{L. Besombes}
\email{lucien.besombes@grenoble.cnrs.fr}
\affiliation{CEA-CNRS group "Nanophysique et
semiconducteurs", Institut N\'eel, CNRS \& Universit\'e
Joseph Fourier, BP 166, F-38042 Grenoble Cedex 9, France}

\date{\today}

\begin{abstract}
We report on the observation of spin dependent optically
dressed states and optical Stark effect on an individual Mn
spin in a semiconductor quantum dot. The vacuum-to-exciton
or the exciton-to-biexciton transitions in a Mn-doped
quantum dot are optically dressed by a strong laser field
and the resulting spectral signature is measured in
photoluminescence. We demonstrate that the energy of any
spin state of a Mn atom can be independently tuned using
the optical Stark effect induced by a control laser. High
resolution spectroscopy reveals a power, polarization and
detuning dependent Autler-Townes splitting of each optical
transition of the Mn-doped quantum dot. This experiment
demonstrates a complete optical resonant control of the
exciton-Mn system.
\end{abstract}

\maketitle

Semiconductor quantum dots (QDs) exhibit discrete excitonic
energy levels with an atom-like light-matter interaction.
Resonant optical excitation of QDs has allowed to observe
the absorption of a single QD \cite{Hogele2004} and the
efficient preparation of the quantum state of a single
confined carrier \cite{Atature2006, Brunner2009} or Mn spin
\cite{LeGall2010}. So far, only a few experiments have
demonstrated the possibility to use a strong continuous
wave laser field to create hybrid matter-field systems and
manipulate QDs states in their solid environment. The
optical dressing of an exciton via the biexciton transition
using absorption spectroscopy \cite{Jundt2008}, the
Autler-Townes effect in the fine structure of the ground
state of a neutral \cite{Xu2007} or charged
\cite{Kroner2008} QD, the Mollow absorption spectrum of an
individual QD \cite{Xu2008} and the emission of optically
dressed exciton-biexciton complex for a QD in a planar
micro-cavity \cite{Muller2008} have been reported. It has
also been demonstrated that the optical Stark effect can be
used to compensate the exchange splitting in anisotropic
QDs to produce entangled photon pairs \cite{Muller2009}.

We show in this letter that the energy of any spin state of
an individual Mn atom embedded in a II-VI semiconductor QD,
can be tuned using the optical Stark effect induced by a
strong laser field. Under resonant excitation, hybrid
light-matter states are created independently for all Mn
spin states. The corresponding Rabi energy measured through
the Autler-Townes splitting can reach 250~$\mu eV$. At low
temperature, the energies that control the dynamics of an
isolated Mn spin in a CdTe QD, like the strain induced
magnetic anisotropy or hyperfine coupling to the nuclei,
are weaker than the observed optical Stark shifts. This
opens a way to control the dynamics of a single Mn spin in
its solid state environment. We also report optical Stark
shifts and optically dressed states of the Mn exchanged
coupled with the exciton or biexciton. Finally, we discuss
a particular situation where two Mn spin states are mixed
by the coupling between bright and dark excitons. In spite
of the Mn spin mixing, we show that an optical manipulation
of individual spin states can be performed.

The sample used in this study is grown on a ZnTe substrate
and contains CdTe QDs. A 6.5 monolayer thick CdTe layer is
deposited at 280$^{\circ}$C by atomic layer epitaxy on a
ZnTe barrier grown by molecular beam epitaxy at
360$^{\circ}$C. The dots are formed by a Tellurium
deposition/desorption process \cite{Maingault2006} and
protected by a 100~nm thick ZnTe top barrier. The QDs are
10-20 nm wide and few nm high. Mn atoms are introduced
during the growth of the CdTe layer. The Mn concentration
is adjusted to optimize the probability to detect one Mn
per dot.

Optical addressing of QDs containing a single magnetic atom
is achieved using micro-spectroscopy techniques. A high
refractive index hemispherical solid immersion lens is
mounted on the surface of the sample to enhance the spatial
resolution and the collection efficiency of single dot
emission in a low-temperature ($T$=5K) scanning optical
microscope \cite{Zwiller2002}. This technique also reduces
the reflected and scattered light at the sample surface
allowing the measurement of spin-flip scattered photons
from a Mn-doped QD \cite{LeGall2010}. Single QD PL is
excited with a continuous wave dye laser tuned to an
excited state of the QD \cite{Glazov2006}. Simultaneously,
a tunable continuous wave single-mode dye ring laser, named
control laser in the following, is used to resonantly
excite the excitonic transitions. The resulting circularly
polarized collected PL is dispersed and filtered by a 1 $m$
double monochromator before being detected by a cooled CCD
camera.

\begin{figure}[hbt]
\includegraphics[width=3.0in]{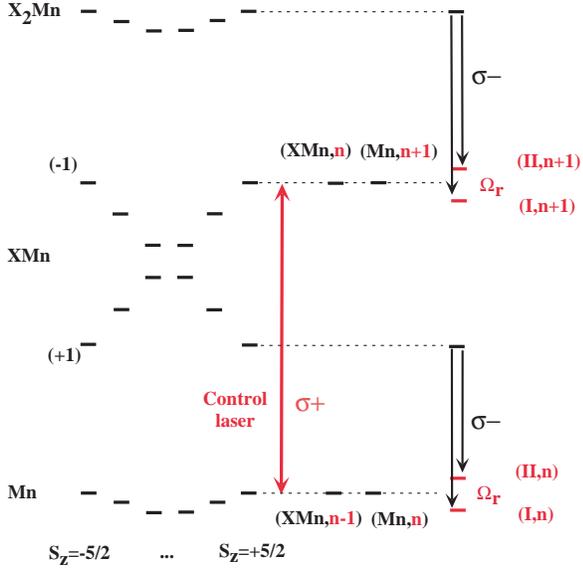}
\caption{Energy scheme of a Mn-doped QD and formation of
light-matter hybrid states by a laser field.  In the
absence of carriers, the Mn fine structure is dominated by
the strained induced magnetic anisotropy which also splits
the biexciton states (X$_{2}$Mn). The bright exciton levels
(X, with kinetic momentum $\pm1$) are split by the exchange
interaction with the Mn (XMn levels). A pump laser tuned to
a QD excited state is used to produce PL of any exciton and
biexciton states. The Rabi splitting, $\hbar\Omega_r$,
induced on the Mn state by the control laser (circularly
polarized $\sigma+$) can be probed in the PL of the exciton
while the splitting of XMn is observed in the PL of the
biexciton.  (I,n), (II,n), (I,n+1) and (II,n+1) are the
optically dressed states produced by the mixing of the
uncoupled states (XMn,n-1), (Mn,n), (XMn,n) and (Mn,n+1)
where $n$ is the number of photons in the control
laser.}\label{fig1}
\end{figure}

When a Mn atom is included in a II-VI semiconductor QD
(CdTe in ZnTe), the spin of the optically created
electron-hole pair (exciton) interacts with the five {\it
d} electrons of the Mn (total spin S=5/2). This leads to a
splitting of the once simple PL spectrum of an individual
QD into six (2S+1) components \cite{Besombes2004}. This
splitting results from the spin structure of the confined
heavy holes which are quantized along the QDs' growth axis
with their spin component taking only the values
J$_z$=$\pm$3/2. In first approximation, the hole-Mn
exchange interaction reduces to an Ising term J$_z$.S$_z$
and shifts the emission energy of the QD, depending on the
relative orientation of the spin state of the Mn (S$_z$)
and hole (J$_z$). As the spin state of the Mn atom
fluctuates during the optical measurements, the six lines
are observed simultaneously in time average PL spectra: The
PL (emission energy and polarization) is a probe of the
spin state of the Mn when the exciton recombines
\cite{Besombes2008}.

\begin{figure}[hbt]
\includegraphics[width=3.5in]{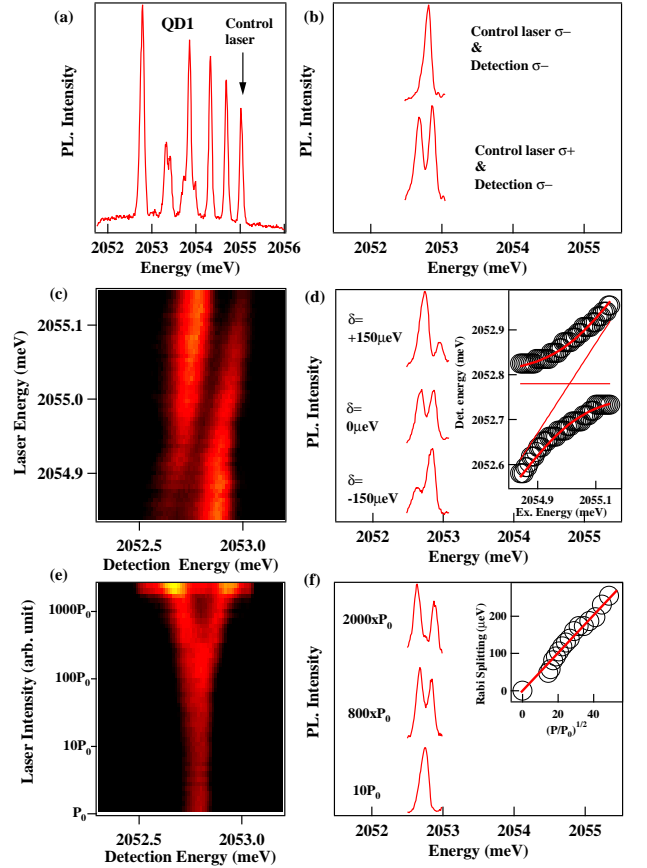}
\caption{Autler-Townes splitting of the emission of
$|-1,+5/2\rangle$ in a Mn-doped QD (QD1) resonantly excited
on $|+1,+5/2\rangle$. (a) shows the non-resonant
photoluminescence of the QD. The intensity map (c) shows
the excitation energy dependence of the Rabi splitting. The
corresponding emission line-shape is presented in (d). The
inset shows the spectral position of the Autler-Townes
doublet as a function of the pump detuning. The fit is
obtained with a Rabi energy $\hbar\Omega_r=180\mu eV$. The
straight lines corresponds to the uncoupled exciton and
laser energy. The excitation intensity dependence of the
Autler-Townes doublet is presented in the intensity map
(e). The corresponding emission line-shape are presented in
(f). The inset shows the evolution of the Rabi splitting as
a function of the square-root of the pump intensity. A
linear increase is observed. (b) presents the circular
polarisation dependence of the Rabi splitting obtained
under resonant excitation.}\label{fig2}
\end{figure}

Only one spin state of the Mn is addressed when a control
laser is circularly polarized ($\sigma\pm$) and tuned on
resonance with an emission line of the exciton-Mn (XMn)
complex. As illustrated in Fig.~\ref{fig1}, the splitting
induced by the control laser tuned to the high energy line
of XMn in $\sigma+$ polarization can be detected in
$\sigma-$ polarization on the low energy line of XMn. This
is the equivalent of the Autler-Townes splitting observed
in atomic physics \cite{Autler1955}. The control laser
field mixes the states with a Mn spin component S$_z$=+5/2
in the presence (XMn) or absence (Mn alone) of the exciton.
At the resonance, the unperturbed states
$|Mn\rangle\otimes|n\rangle$ and
$|XMn\rangle\otimes|n-1\rangle$ can be dressed into pairs
of hybrid matter-field states $|I,n\rangle$ and
$|II,n\rangle$ where $|n\rangle$ is a n-photons sates of
the control laser (see Fig.~\ref{fig1}). These states can
be written as \cite{Boyle2009}:
 \begin{eqnarray*}
|I,n\rangle=c|Mn\rangle\otimes|n\rangle-s|XMn\rangle\otimes|n-1\rangle\\
|II,n\rangle=s|Mn\rangle\otimes|n\rangle+c|XMn\rangle\otimes|n-1\rangle
\end{eqnarray*}
with corresponding energies
$E_{\pm}=\frac{\hbar}{2}(\omega_c+\omega_0)\pm\frac{\hbar}{2}\Omega_r^{\prime}$.
Here, \noindent
$c=\sqrt{\frac{1}{2}(1-\frac{\delta}{\Omega^{\prime}})}$
and
$s=\sqrt{\frac{1}{2}(1+\frac{\delta}{\Omega^{\prime}})}$.
$\delta=\omega_c-\omega_0$ is the laser detuning with
$\omega_0$ the resonance frequency of the unperturbed
transition and $\omega_c$ the frequency of the control
laser.
$\hbar\Omega_r^{\prime}=\hbar\sqrt{\Omega_r^2+\delta^2}$
defines the energy splitting of the dressed states where
$\Omega_r=\mathcal{P}\mathcal{E}/\hbar$ is the Rabi
frequency with $\mathcal{P}$ the dipolar moment of the QD
transition and $\mathcal{E}$ the amplitude of the electric
field of the control laser. A power dependent Autler-Townes
type splitting is then expected for all transitions that
share such an optically dressed state
\cite{Autler1955,Mollow1972}.

Experimental data corresponding to a control laser tuned on
$|+1,+5/2\rangle$ and the observation of an Autler-Townes
splitting in the PL of the state $|-1,+5/2\rangle$ are
presented in Fig.~\ref{fig2}. Particular care is given to
the effect of the detuning of the control laser from the
XMn resonance (Fig.~\ref{fig2}(c) and \ref{fig2}(d)) and
its intensity (Fig.~\ref{fig2}(e) and \ref{fig2}(f)). At
large laser detuning, the optically active transitions
asymptotically approach the original excitonic transitions
where the remaining energy offset is the optical Stark
shift. At the resonance, an anti-crossing is observed
showing that the strong coupling between the laser field
and the exciton creates hybrid light-matter states. As
presented in the inset of Fig.~\ref{fig2}(d), a good
agreement with the simple dressed atom model is obtain with
a Rabi energy of $\hbar\Omega_r$=180$\mu$eV. On resonance,
the emission from the $|-1,+5/2\rangle$ state splits into a
doublet when the power of the control laser is increased,
as expected from the Autler-Townes model. The splitting is
plotted as a function of the square root of the control
laser intensity in Fig.~\ref{fig2}(f), showing that the
splitting linearly depends on the laser field strength. A
Rabi splitting larger than 250$\mu$eV is obtained at high
excitation intensity. It is worth noting that these energy
shifts can be easily larger than the magnetic anisotropy of
an isolated Mn spin created by the strain in the QD plane
($\approx$40$\mu$eV)
\cite{Qazzaz1995,Goryca2009,LeGall2009}. This optical
tuning of the fine structure may lead to a control of the
coherent dynamics of the isolated Mn spin.

The high energy transition of the XMn complex is twice
degenerated. The corresponding optical transitions differ
by the polarization of the absorbed or emitted photons. The
polarization dependence of the laser induced splitting
shown in Fig.~\ref{fig2}(b) confirms the Mn spin
selectivity of the strong coupling with the laser field:
$\sigma+$ photons couple with the state $|+5/2\rangle$ of
the Mn to create two hybrid light-matter states while no
splitting of the $\sigma-$ PL line is observed with
$\sigma-$ control photons.

\begin{figure}[hbt]
\includegraphics[width=3.3in]{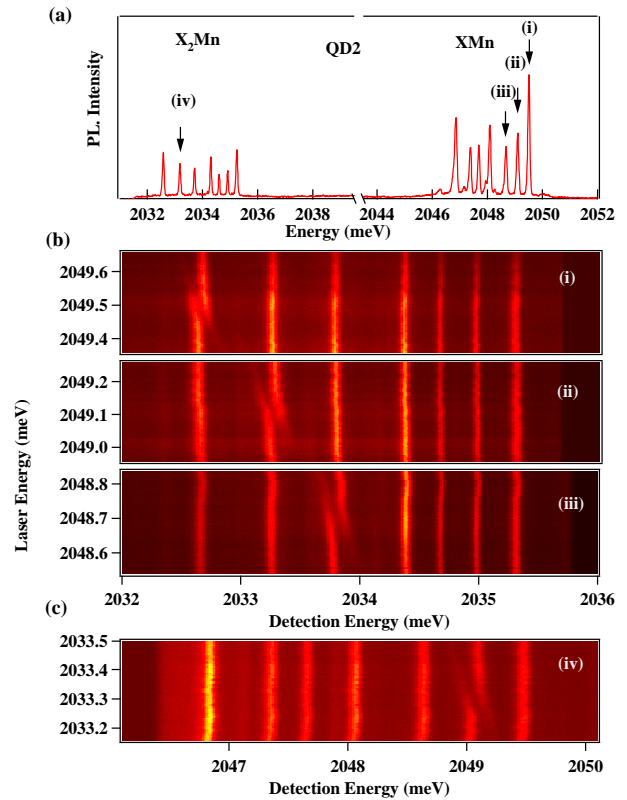}
\caption{(a) PL of the exciton and biexciton in a Mn-doped
QD (QD2). (b) Autler-Townes splitting of the exciton in QD2
detected on the biexciton PL under resonant excitation of
the ground-to-exciton transition for the spin state of the
Mn S$_z$=+5/2 (i), S$_z$=+3/2 (ii) and S$_z$=+1/2 (ii)
(arrows in the PL of QD2). (c) Emission of the exciton for
a dressed exciton-to-biexciton transition. The excitation
is tuned around the state S$_z$=+3/2 of the biexciton
(iv).}\label{fig3}
\end{figure}

The strong coupling with the control laser is also observed
in optical transitions that involve the biexciton exchanged
coupled to a single Mn ($X_{2}$Mn). This is illustrated in
Fig.~\ref{fig3} in the case of successive resonant
excitations on the XMn levels with a Mn spin state
S$_z$=+1/2, +3/2 and +5/2. In these cases, the
recombinaison of $X_{2}$Mn probes the laser induced
splitting of XMn for a given spin states of the Mn. It is
shown here that any XMn transition, and consequently any Mn
spin state, can be optically shifted by a control laser
tuned on resonance. As illustrated in Fig.~\ref{fig3}(c),
by coherently driving the $X_{2}$Mn-to-XMn transition, one
can also tune the energy of any state of the XMn complex.
This set of experiment demonstrates that a complete optical
control of the exciton-Mn system is possible.

\begin{figure}[hbt]
\includegraphics[width=3.3in]{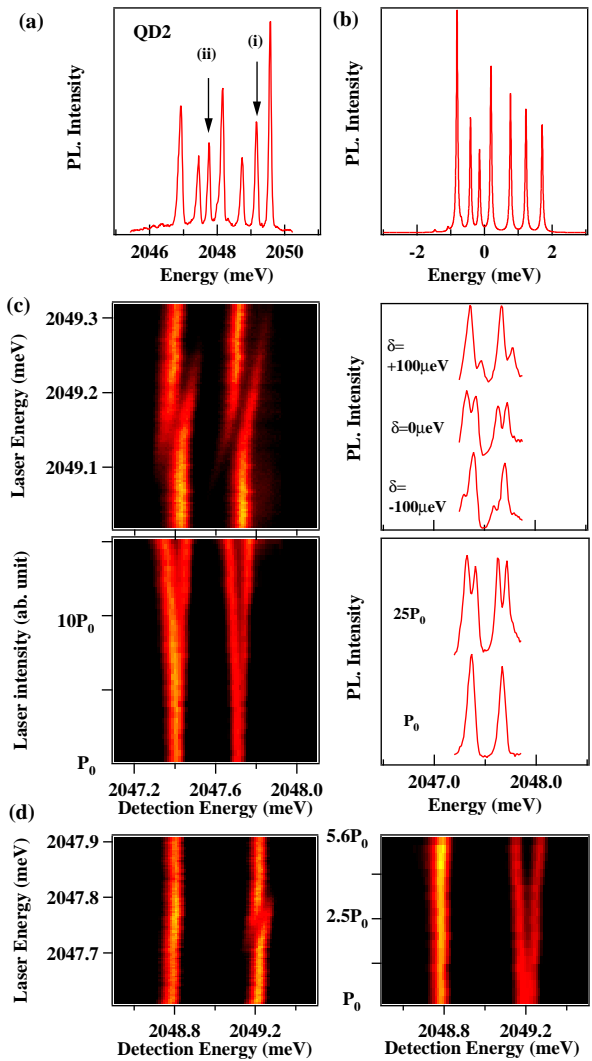}
\caption{Rabi splitting obtained on mixed bright-dark
exciton states in QD2. (a) presents the PL of QD2. The
intensity of the lines is influenced by the absorption
selectivity of the excited state (see ref.
\cite{Glazov2006}). The corresponding theoretical emission
spectra is presented in (b). It is calculated with
$J_{eMn}$=-~0.095 meV, $J_{hMn}$=0.3 meV, $J_{eh}$=-0,6 meV
and $\epsilon_{vbm}=0.1$, $\epsilon_{vbm}=0.1$,
$\delta_2$=0.05 meV and T=25K (see ref. \cite{Leger2007}
for a detail of the model). (c) presents the laser detuning
dependence and the excitation power dependence of the Rabi
splitting obtained on mixed bright-dark exciton under
excitation on (i). (d) presents the detuning and excitation
intensity dependence measured on the high energy
transitions associated with the Mn spin state S$_z$=+1/2
and S$_z$=+3/2 under excitation on the mixed dark-bright
exciton (ii).}\label{fig4}
\end{figure}

It is also demonstrated here that the use of a resonant
strong laser field allows to individually address any spin
state of the Mn even if they are coupled by the exciton
through the valence band mixing (VBM) \cite{Leger2007}. The
particular situation where the Mn spin states
$|+1/2\rangle$ and $|+3/2\rangle$ are significantly mixed
is presented in Fig.~\ref{fig4}. The spectrum of this QD
(Fig.~\ref{fig4}(a)) differs clearly from the one expected
in the pure heavy-hole approximation and presents seven
peaks. Such PL spectrum appears when the high energy lines
of the dark exciton states overlap the low energy lines of
the bright exciton states. When dark and bright exciton
states are close in energy in the heavy hole approximation,
simultaneous hole-Mn spin flips allowed by the strain
induced VBM, mixes the dark and bright states: from one
dark and one bright state, one gets two split states with a
bright component \cite{Leger2007}. This is the case for
QD2: the VBM couples $|-1,+3/2\rangle$ with
$|+2,+1/2\rangle$ and the new eigenstates share the
oscillator strength of the bright state $|-1,+3/2\rangle$.
The two lines on the right of the low energy state can be
attributed to the bright part of mixed bright-dark
excitons. This attribution is confirmed by the calculation
of the energy levels presented in Fig.~\ref{fig4}(b)
\cite{JFR2006,Leger2007}.

As shown in Fig.~\ref{fig4}(c) and \ref{fig4}(d), it is
possible to optically address selectively one state (and
one only) of the Mn spin in the mixed bright-dark XMn
levels. When the $\sigma+$ control laser is tuned on the
state $|+1,+3/2\rangle$, a Autler-Townes splitting is
observed in $\sigma-$ polarization for both components of
the emission of the dark-bright excitonic complexes
(Fig.\ref{fig4}(c)). This arises from the control laser
induced splitting of their common final state with a Mn
spin S$_z=+3/2$. With a resonant excitation on the mixed
bright-dark states, only the states which share a
S$_z=+3/2$ Mn spin are split: the dark part with Mn spin
S$_z=+1/2$ is not affected. This is demonstrated in
Fig.\ref{fig4}(d). The $\sigma-$ control laser splits the
Mn state $|+3/2\rangle$ which leads to the doublet
formation in the $\sigma+$ PL from the state
$|+1,+3/2\rangle$ (PL around 2049.2 meV) while no splitting
of the state $|+1,+1/2\rangle$ is observed. The Mn spin
state S$_z=+1/2$ is not affected by the control laser tuned
on the dark-bright mixed exciton. This experiment suggests
the possibility to optically control the exciton induced
coupling between two spin states of the Mn atom, an
important step towards coding quantum information on an
individual magnetic atom \cite{Reiter2009}.

In summary, we have demonstrated that the ground, exciton
and biexciton states in a Mn-doped QD can be coherently
manipulated by applying a strong resonant laser field on
the optical transitions. At the resonance, hybrid
matter-field states are created that should significantly
influence the Mn spin dynamics. Our results demonstrate
that even under strong optical field, the transition in a
Mn-doped QD behaves like isolated two-level quantum systems
well described by the dressed atom picture. In the ground
state, the laser induced shift of the Mn spin could be used
for a fast optical manipulation of the spin degree of
freedom of the Mn atom.

This work is supported by the French ANR contract QuAMOS,
Fondation Nanoscience (Grenoble) and EU ITN project
Spin-Optronics.

\end{document}